\PassOptionsToPackage{dvipsnames}{xcolor}
\documentclass[%
 reprint,
 superscriptaddress,
 twocolumn,
 amsmath,amssymb,
 aps,
prx,
floatfix,
showkeys,
]{revtex4-2}
\setcitestyle{super,sort&compress}
\usepackage{graphicx}
\usepackage{dcolumn}
\usepackage{bm}
\usepackage{siunitx}
\usepackage{orcidlink}

\usepackage{import}
\usepackage[nameinlink]{cleveref}

\usepackage{microtype}

\DeclareRobustCommand{\fakeSC}[1]{%
  {\scalebox{0.8}[0.8]{\MakeUppercase{#1}}}%
}

\usepackage{algorithm}[compatible]
\usepackage{algpseudocode}
\algrenewcommand\algorithmicindent{0.5em}

\usepackage[columnwise]{lineno}

\usepackage{enumitem}
\usepackage{amsthm}

\usepackage{pdfpages}
\setboolean{@twoside}{false}
\makeatletter
\AtBeginDocument{\let\LS@rot\@undefined}

\usepackage{hyperref}

\hypersetup{
    colorlinks=true,
    linkcolor=Magenta,
    citecolor=ForestGreen,
    urlcolor=Cerulean
}

\makeatletter

\renewcommand{\section}{%
  \@startsection
    {section}%
    {1}%
    {\z@}%
    {-0.6cm}%
    {0.2cm}%
    {%
      \normalfont\small\bfseries
      \raggedright
    }%
}

\renewcommand{\subsection}{%
  \@startsection
    {subsection}%
    {2}%
    {\z@}%
    {-0.6cm}%
    {0.12cm}%
    {%
      \normalfont\small\bfseries
      \raggedright
    }%
}

\renewcommand{\subsubsection}{%
  \@startsection
    {subsubsection}%
    {3}%
    {\z@}%
    {-0.4cm}%
    {0.06cm}%
    {%
      \normalfont\small\itshape
      \raggedright
    }%
}

\makeatother

\begin{document}
\title{GA\fakeSC{nn}ET: A principled approach to automatically\\annotating charge stability diagrams}

\author{Brian Weber}
\affiliation{Intelligent Geometries, LLC, Lake Frederick, VA 22630, USA}
\affiliation{Technology Test and Evaluation Division, National Institute of Standards and Technology,\\Gaithersburg, MD 20899, USA}
\affiliation{Joint Center for Quantum Information and Computer Science,
University of Maryland, College Park, MD 20742, USA}

\author{Justyna P. Zwolak\orcidlink{0000-0002-2286-3208}}
\email{jpzwolak@nist.gov}
\affiliation{Technology Test and Evaluation Division, National Institute of Standards and Technology,\\Gaithersburg, MD 20899, USA}
\affiliation{Joint Center for Quantum Information and Computer Science,
University of Maryland, College Park, MD 20742, USA}
\affiliation{Department of Physics, University of Maryland, College Park, MD 20742, USA}

\begin{abstract}
Gate-defined semiconductor quantum-dot (QD) arrays are a promising platform for quantum computing.
However, their large configuration spaces and inherent measurement noise make device tuning time consuming and difficult to scale.
Although machine-learning (ML) approaches for automated tuning have proliferated rapidly, a fair comparison of methods remains limited by inconsistent metrics and a lack of standardized labeled datasets. 
Here, we introduce Geometry-Informed ANNotator for Electrostatic Topologies (\textsc{GAnnET}), a physics- and geometry-informed algorithm that automatically generates probabilistic state labels for charge stability diagrams (CSDs) of double-QD systems. 
\textsc{GAnnET} constructs local geometric models of CSD structure and combines them into a global state map with calibrated confidence, eliminating the need for manual annotation. 
We validate \textsc{GAnnET} on simulated and experimental CSDs and demonstrate consistent labeling suitable for ML training and benchmarking. 
By enabling reproducible annotation of experimental data, \textsc{GAnnET} supports community benchmarking and scalable automated QD operation.
\end{abstract}


\maketitle
\section*{Introduction}
\label{sec:introduction}
Semiconductor-based quantum-dot (QD) arrays, in which charge carriers are confined in gate-defined potential wells and used to encode and carry quantum information, offer the selectivity and connectivity needed for large-scale quantum computing.~\cite{Zwanenburg13-SQE, Burkard21-SSQ}
At the same time, realizing this promise at scale remains challenging as device behavior remains sensitive to fabrication imperfections, electrostatic disorder, and environmental fluctuations.~\cite{Neyens24-PQW, Massai24-ITC}
Operating QD devices requires careful adjustment of multiple gate voltages to define charge occupancy and set tunnel coupling between neighboring QDs (for two-qubit interactions) and between QDs and reservoirs (for initialization and readout), which is a nontrivial, time-consuming process.
Tuning state-of-the-art QD devices already involves adjusting tens of electrostatic and dynamical control parameters that must be carefully calibrated to reach the single-charge-carrier regime and achieve high-fidelity qubit operation.~\cite{Wang24-OQP, George24-QAM}
As the number of QDs increases, the relevant configuration space grows rapidly, making manual and heuristic-based control increasingly time consuming and error prone. 

These challenges have motivated substantial progress in automating the tuning workflow, including approaches that begin at device turn-on~\cite{Darulova19-ATQ, Zubchenko24-ABQ, Kovach24-BATIS} or focus on charge tuning, targeted calibration, and in situ optimization tasks.~\cite{Baart16-CAT, Durrer19-ATQ, Kalantre17-MLD, Zwolak20-AQD, Zwolak21-RBI, Ziegler22-TAR, Czischek21-MNA, Lapointe-Major19-ATQ, Ziegler23-AEC, Benestad25-AIO}
Early demonstrations often relied on conventional algorithmic tools such as regression, pattern matching, and feedback control.~\cite{Baart16-CAT, Botzem18-TSD, Lapointe-Major19-ATQ}
More recently, the field has increasingly adopted supervised and unsupervised machine learning (ML), often combined with device-physics priors, to accelerate and generalize tuning decisions.~\cite{Zwolak21-AAQ}

Despite rapid methodological progress, the QD community still lacks a consistent approach for comparing autotuning methods across various devices, noise conditions, and operating regimes.
ML-based approaches, in particular, require large labeled datasets for both training and evaluation, yet available labels are often manual, qualitative, and challenging to reproduce.
Moreover, the definitions and interpretations of reported performance metrics can vary widely, complicating the benchmarking of ML methods against conventional approaches or against one another.~\cite{Zwolak21-AAQ}
Establishing standardized datasets and evaluation protocols is, therefore, a crucial step toward reliable and scalable QD autotuning.~\cite{Zwolak23-DNC}

A central obstacle is the availability of datasets for development and testing.
Existing openly available resources generally fall into two broad categories: datasets generated using simulations of charge-stability measurements, often augmented with synthetic noise to mimic measurement variability and device imperfections,~\cite{Buterakos25-QDF, Buterakos25-QDFcode, vanStraaten24-QAr, vanStraaten24-QArcode, Krzywda25-QDa, Krzywda25-QDacode, Gualtieri25-QDsim, Gualtieri25-QDsimcode, Hader24-SCS} and experimentally acquired datasets that require manual annotation.~\cite{qf-data, DCSD-data}
Each of these categories comes with its own challenges. 
Simulated data may miss important experimental artifacts and device-specific imperfections, whereas manually labeled experimental data can be subject to inconsistency and error.
Moreover, with a few exceptions, such datasets have not been released publicly.
Analogous to the role played by MNIST~\cite{MNIST-paper} and CIFAR~\cite{CIFAR-report} in computer vision, broadly adopted benchmark datasets are necessary to enable a systematic comparison of methods and clearer measurement of progress.

The so-called \textit{charge stability diagrams} (CSDs)\,---\,two-dimensional (2D) scans of a device response versus pairs of gate voltages that compactly encode where charge transitions occur and how charge configurations evolve across control space\,---\,are a primary diagnostic for QD device operation.
The transition lines in CSDs correspond to the movement of charge carriers to and from a QD via nearby reservoirs or to the transfer of charge carriers between neighboring QDs. 
The resulting geometric arrangement and connectivity of charge-state regions encode the device's underlying \textit{electrostatic topology} in measurement space.
The orientation of charge transitions encodes relative lever arms between gates and QDs, while line spacings determine charging energies and mutual capacitances (up to an overall energy scale), enabling an effective constant-capacitance characterization that provides feedback on gate layout and heterostructure design.~\cite{Wiel02-DQD, vanStraaten24-QAr, vanStraaten24-QArcode, Gualtieri25-QDsim, Gualtieri25-QDsimcode}
These extracted quantities are also central to downstream tune-up tasks such as virtual-gate construction and robust control protocols.~\cite{Oakes20-AVE, Ziegler23-AEC, Rao24-MAViS, Rao25-TFS} 
As QD systems scale, rapid and reliable extraction of CSD features becomes a prerequisite for autonomous control: MAViS, for example, uses CSD-derived features to autonomously construct multi-layer virtualization in two-dimensional QD arrays.~\cite{Rao24-MAViS}
Recent works use CSDs as the substrate for automated electrostatic characterization across single- and bilayer heterostructures~\cite{Losert26-AEC} and for time-domain calibration and stabilization by tracking the motion of charge-transition networks in repeatedly acquired CSDs.~\cite{Rao25-TFS}
However, systematic benchmarking and robust generalization require substantially larger standardized collections of experimental CSDs spanning different materials, designs, and operating conditions.~\cite{Zwolak23-DNC}

\begin{figure}[t]
    \centering
    \includegraphics[width=\linewidth]{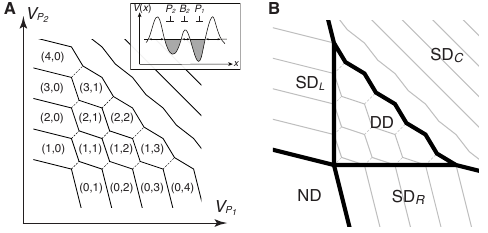}
    \caption{\textbf{Schematic CSD and corresponding global state map.}
    (A) Schematic CSD showing the irregular polygonal tiling of the $(V_{P_1}, V_{P_2})$ configuration space.
    The $(m,n)$ labels denote the ground-state charge configuration of the left and right QDs.
    The inset shows a schematic potential landscape in the DD state.
    (B) Corresponding global \textit{state map} partitioning of the $(V_{P_1},V_{P_2})$ configuration space into five domains: ND, SD$_L$, SD$_C$, SD$_R$, and DD.
    }
    \label{fig:tiling-model}
\end{figure}

To facilitate systematic processing of experimentally acquired 2D CSDs, we introduce \textit{Geometry-Informed ANNotator for Electrostatic Topologies} (\textsc{GAnnET}), a tool for automated, unbiased analysis and labeling of CSDs.
\textsc{GAnnET} combines geometric methods and physical insights to decompose a CSD into physically meaningful \textit{domains} in which the QD system exhibits consistent behavior, thereby recovering the electrostatic topology encoded in the CSD.
Because the analysis stages in \textsc{GAnnET} are theoretically motivated and grounded in mathematical structure, the resulting annotations are reproducible and suitable for downstream ML development, evaluation, and benchmarking.
\textsc{GAnnET} is designed to assist in creating a standardized database of experimental CSDs, helping bridge the gap between algorithm design and practical experimentation.

\textsc{GAnnET} is robust to variability in device parameters such as gate design or interdot coupling strength.
In this work, we validate \textsc{GAnnET} on a publicly available dataset of double-QD (DQD) CSDs from two undoped accumulation-mode QD devices.~\cite{qf-data}
However, the underlying geometric and statistical framework is not specific to this particular device realization or to double-dot (DD) systems alone.
Because \textsc{GAnnET} is based on a polytopal description of charge-state regions, it can, in principle, be configured to recognize more general polytopal domains and to operate in higher-dimensional measurement spaces.
In addition, the simulated ensemble used here spans deliberately diverse device realizations rather than a single narrowly tuned model instance.
Broader experimental validation across additional device geometries, material platforms, and sensing modalities remains an important direction for future work.
In addition to producing state labels, the geometric structure recovered by \textsc{GAnnET} yields interpretable features that can support diagnostic reasoning and identification of failure modes.
These richer geometric features, such as apertures and other local polygonal attributes, are not quantitatively benchmarked in the present work but could support future identification of finer-grained regions of interest relevant for tuning.

\section*{Results}
\label{sec:results}
The initial goal of tuning semiconductor QD systems for quantum computing is to electrostatically confine charge carriers into controllable potential wells using gate electrodes.
In a typical gate-defined device, screening gates shape current channels and suppress the formation of unintended charge islands, plunger gates primarily set the local electrochemical potentials that control dot occupancy, and barrier gates tune the tunnel barriers between wells and between wells and reservoirs (see the inset in Figure~\ref{fig:tiling-model}A).
When the potential wells are sufficiently isolated and occupied, they form QDs.

\begin{figure*}[t]
    \centering
    \includegraphics{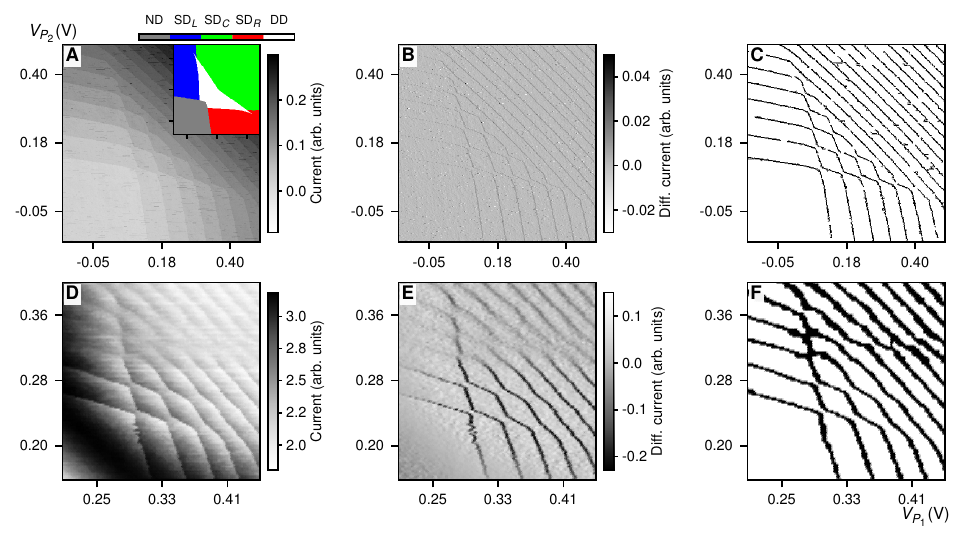}
    \caption{\textbf{Preprocessing of the binary transition mask for a sample simulated (top) and experimental (bottom) CSD.}
    (A) Simulated 2D measurement generated with the QD simulator~\cite{Zwolak18-QLD}.
    The inset shows the corresponding ground-truth state labels from the simulation.
    (B) Numerical derivative of (A), highlighting charge-transition features and revealing the irregular polytope tiling.
    (C) Binary transition mask derived from (A), which serves as the required \textsc{GAnnET} input.
    (D) Experimental 2D measurement from the QFlow 2.0 dataset~\cite{qf-data}.
    (E) Numerical derivative and (F) binary transition mask derived from (D).
    }
    \label{fig:masking-sample}
\end{figure*}

\subsection*{Electrostatic topologies in charge stability diagrams}
\label{ssec:elec_top}
In a typical DQD device, five gates\,---\,two plungers and three barriers\,---\,are used to define and control the two QDs.
If the barrier gate voltages are held fixed, the plunger voltages $V_{P_1}$ and $V_{P_2}$ span a two-dimensional \emph{configuration space} in which the device exhibits distinct charge configurations.
The discrete nature of charge yields a CSD that divides the configuration space into regions corresponding to stable ground-state charge configurations.
Under standard assumptions of weak tunnel coupling and approximately constant capacitances, the constant-interaction model predicts that these regions are (approximately) convex polytopes in the plunger-gate plane.~\cite{Beenakker91-COC, Schroer07-ETQ, Wiel02-DQD}
A schematic example of such an irregular tiling, together with the charge configuration labels for each tile, is shown in Figure~\ref{fig:tiling-model}A.

At a coarse level, the DQD configuration space can be divided into five \emph{global state domains}: DD; left, central, and right single dot (SD$_L$, SD$_C$, and SD$_R$, respectively); and no dot (ND).
Within a given domain, the polytope geometry, including typical sizes, orientations, and facet structure, reflects the device’s qualitative electrostatic behavior (e.g., whether charge localizes on the left versus the right QD).
Polytopes with similar characteristics cluster together, allowing the subdivision of the CSDs into the distinct global state domains.
Moreover, transitions between global state domains occur in a monotonic order, meaning that polytopes with similar geometric characteristics cluster in contiguous regions of the configuration plane.
This observation motivates an algorithm that infers global domains by extracting and aggregating local geometric structure.
A schematic domain partition is shown in Figure~\ref{fig:tiling-model}B.

\subsection*{Datasets overview}
\label{ssec:data}
\textsc{GAnnET} is validated on both simulated and experimental CSDs.
Simulated data are generated using QDFlow,~\cite{Zwolak18-QLD, Buterakos25-QDF, Buterakos25-QDFcode} a physics-based QD simulator that produces 2D measurements analogous to those acquired in the laboratory.
Figure~\ref{fig:masking-sample}A shows an example simulated CSD, with the associated ground-truth global state labels shown in the inset. 
The corresponding numerical derivative, emphasizing charge-transition features and revealing the characteristic polygonal tiling of the plunger-gate plane, is shown in Figure~\ref {fig:masking-sample}B.
Figures~\ref{fig:masking-sample}D and \ref{fig:masking-sample}E give an example of an experimentally acquired CSD from the QFlow 2.0 dataset~\cite{qf-data} and its numerical derivative, respectively. 

The final \textit{binary transition mask} that \textsc{GAnnET} operates on to recover the underlying tiling structure and to infer the global state-domain partition of the configuration space, shown in Figures~\ref{fig:masking-sample}C and \ref{fig:masking-sample}F, respectively, is derived through a combination of thresholding and edge detection applied to the numerical derivatives of CSDs (see \hyperref[sec:methods]{methods}).
Across both simulated and experimental CSDs, the two scan axes correspond to experimentally tunable control parameters (here, the plunger gate voltages $V_{P_1}$ and $V_{P_2}$; see Figure~\ref{fig:tiling-model}A).
The resulting CSD patterns arise from discrete changes in charge-carrier occupation, which are visible as charge-transition features.

\subsection*{Overview of the GA\fakeSC{nn}ET workflow}
\label{ssec:alg_overview}
The \textsc{GAnnET} algorithm has three distinct phases.
\begin{description}[itemsep=-2pt,topsep=6pt,leftmargin=20pt]
    \item[\textit{Phase 1}] \textit{\textbf{Model building}}\,---\,Collecting the information necessary to build discrete polygonal models using the \textit{fingerprinting} method.~\cite{Zwolak20-RBC, Weber21-TBR}
    \item[\textit{Phase 2}] \textit{\textbf{Statistical inference}}\,---\,Using the gross features of the polygons, i.e., orientation, interior angles, number of edges, etc., to group the polygons into classes.
    Because noise effects can distort polygon models in unpredictable ways, this phase also makes noisiness and reliability determinations to distinguish reliable from unreliable models.
    \item[\textit{Phase 3}] \textit{\textbf{Global state determination}}\,---\,Dividing the CSD into global state domains by first using the grouping results from the statistical inference phase and then creating models of the global domains based on the underlying physical principles.
\end{description}

As the interdot coupling between the QDs increases,~\cite{Wiel02-DQD} the DD state grades into the SD$_L$, SD$_C$, or SD$_R$ state without any sharp boundary. 
To accommodate this, the \textsc{GAnnET} algorithm examines each polygon individually and applies explicit external rules based on charge carriers confinement physics~\cite{Wiel02-DQD} to assign a probability that a given polygon belongs to the DD or SD$_C$ state.
Given that the fingerprinting method is error prone in the presence of noise, the \textsc{GAnnET} relies on high levels of measurement redundancy, which is easy to accommodate in the offline setting, i.e., when used to analyze pre-measured scans as opposed to data acquired in situ in a real-time setting, along with certain statistical methods to obtain reliable state labeling.
An overview of the \textsc{GAnnET} workflow is presented in Algorithm~\ref{alg:GAnnET-flow}.
The procedural flow of the \textsc{GAnnET} algorithm, in the form of pseudocode, is provided in the supplemental information.

\subsection*{Geometry-informed polygon modeling}
\label{ssec:poly_model}
The first step in \textsc{GAnnET} is creating polygon models from the binary transition mask.
The primary tool here is fingerprinting.
If $x_o$ is an observation point sampled within a 2D CSD, the \textit{point fingerprint} $\mathcal{F}_{x_o}$ at an observation point $x_o$ is defined as a list of weighted distances from that point to the nearest charge transition line along evenly spaced one-dimensional (1D) measurements called \textit{rays}.
The idea of using fingerprints to classify simple high-dimensional geometrical structures was first proposed in the context of cost-effective calibration of QD devices~\cite{Zwolak20-RBC, Weber21-TBR} and subsequently implemented in autotuning workflows that use learned or geometry-based representations of charge-transition structures to guide automated analysis and optimization.~\cite{Zwolak21-RBI, Krause22-ECP, Krause22-LCD}

While qualitative information about the boundaries defining the polytopes suffices for ML-driven classification, capturing the smooth transition between states for labeling 2D CSDs requires full modeling of the structures of interest. 
In addition to fingerprints, the model-building module requires information about the orientation of each ray to determine the positions of the \textit{terminal points}, i.e., points in the 2D configuration space where rays cross transition lines.
We call the fingerprint combined with a vector of terminal points and ray orientations an \textit{extended point fingerprint}.

\begin{figure}[t]
\begin{algorithm}[H]
\caption{Overview of the \textsc{GAnnET} workflow}
\label{alg:GAnnET-flow}
\begin{algorithmic}[1]
\Procedure{GAnnET}{CSD}
    \State Build and centralize local fingerprints
    \State Construct local polygon models
    \State Filter unreliable models and infer dominant orientations
    \State Build a coarse domain map
    \State Refine the central region into SD$_C$ and $\mathrm{DD}$
    \State Construct the final probabilistic five-state map
    \State \Return final probabilistic five-state map
\EndProcedure
\end{algorithmic}
\end{algorithm}
\end{figure}

\subsection*{Localizing observation points}
\begin{figure*}[t]
    \centering
    \includegraphics[width=0.9\linewidth]{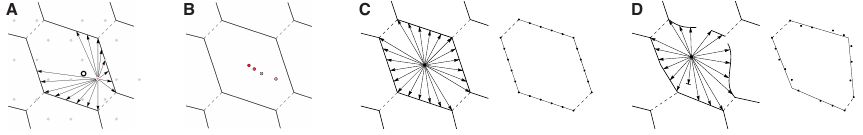}
    \caption{\textbf{The polygon modeling.}
    (A) A preview of a fingerprint measurement from a sample point (shown in pink) from the initial grid of observation points.
    A preview of the initial grid of sampled points is shown in gray.
    (B) A series of consecutive observation points selected by the model-building module, with the intensity of the color indicating the progress of the algorithm.
    (C) Model creation for a nearly ideal polygon (right) and the resulting polygon (left). 
    (D) A corrected set of rays for a noisy model of a polygon (right) and the resulting polygon model (left).
    }
    \label{fig:model-building}
\end{figure*}

The model-building process starts with selecting observation points from which to measure the extended fingerprints.
Ideally, each observation point should be located at the center of a polytopal domain.
This desired arrangement of the centralized observation points is achieved through an iterative process.
Starting with a selection of initial observation points on a dense hexagonal grid with points spaced every four pixels, an extended fingerprint is measured at each point and then fitted to rough star-shaped polygon models.~\cite{SingerThorpe}
The centers of mass of the resulting models become new observation points (see Figure~\ref{fig:model-building}A).
Iterating, this causes the points to cluster at the centers of polygons in regions where convex polygons are captured in the binary transition mask (see Figure~\ref{fig:model-building}B).

In regions where the transition map has parallel or near-parallel lines, observation points do not cluster at the centers but rather along the median lines.
A simple process of pruning reduces the initially very large number of points to a relatively few: if the distance between two observation points is less than the interior radius of either of the polygons they define, the two points are combined into a single point at the midpoint between them.
The process continues until the locations of the observation points stabilize.

\subsection*{Fingerprinting charge stability diagrams}
Having found properly distributed observation points, the next step in the \textsc{GAnnET} algorithm is to measure extended fingerprints at these points and build the final convex polygon models from each fingerprint (see Figure~\ref{fig:model-building}C).
While this is a straightforward process in noise-free settings, data binarization may lead to three types of errors.
These are (1) noise artifacts misidentified as indicators of charge transitions, (2) transition lines with gaps when the thresholding detects no charge transition where a transition should be present, and (3) stochastic imprecision in transition line locations.

The first two types of errors may severely hinder the model-building attempt, and thus, the algorithm must identify and correct them.
In brief, some rays will strike a noise pixel before a transition line, resulting in a too-short ray, while others might miss a transition that should be present, resulting in a too-long ray.
A dense sampling of rays makes it relatively easy to identify individual or small clusters of ray anomalies, and additional checks for deviations from convexity allow the identification of larger groups of anomalous rays.
All anomalous rays are removed from the analysis prior to modeling (see Figure~\ref{fig:model-building}D).

The imprecise positioning of the charge transitions is addressed by using best-fit techniques to model extended fingerprint terminal points as line segments.
While Gaussian best-fit methods are often used in such situations, modeling with a Gaussian cost function is unstable and, even in reasonable cases, may produce poor fits to the data; see, e.g., Simpson's paradox.~\cite{Simpson51-IIC, Yule03-TAA, Pearson99-GRS}
Instead, we use the more stable normalized $\ell^2$ Hausdorff distance (see \hyperref[sec:methods]{methods}) to measure the fidelity of line segments to groupings of terminal points.~\cite{Evans15-MTP}

\subsection*{Fitting and modeling}
\label{sssec:fit_and_model}
The next phase of the geometry-informed polygon modeling involves fitting each extended fingerprint to a model.
There are four kinds of polygons possible in a DQD CSD: 
\begin{itemize}[itemsep=-2pt,topsep=2pt]
    \item hexagons with two missing edges (representing mainly the DD state),
    \item pentagons with one missing edge (representing the SD$_L$ and SD$_R$ states),
    \item quadrilaterals (representing the SD$_C$ state), and
    \item triangles (representing the ND state).
\end{itemize} 
Any missing edge in a model, should any exist, is called an aperture.
In real systems, the aperture might be visible. 
Although such features are physically interpretable and may be useful for identifying finer-grained tuning targets, the present study evaluates only the resulting domain decomposition and does not separately benchmark feature-level localization accuracy.
This is especially true when the sensor dot is positioned asymmetrically between the two sides of the DD being measured.
This choice of geometric models is informed by the constant-interaction model for DQD systems.~\cite{Wiel02-DQD}

Each model is defined by a finite number of parameters, including its vertex points and the presence of apertures.
Every choice of parameters determines a new polygonal model, which is then discretized.
Letting this discretization be the set $A$, and the fingerprint endpoints be the set $B$, we evaluate the normalized $\ell^2$ Hausdorff distance $\rho_H(A,B)$.
Interpreting this as a cost function, we then search for model parameters that minimize it (see \hyperref[sec:methods]{methods}).

\begin{figure*}[t]
    \centering
    \includegraphics[width=0.92\linewidth]{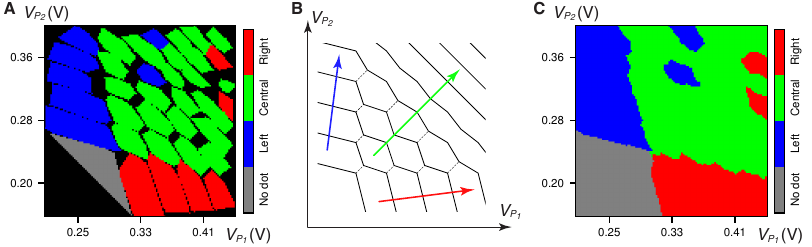}
    \caption{
    \textbf{The depiction of the dominant directions in CSD.}
    (A) Fingerprint-modeled polygons for the CSD shown in Figure~\ref{fig:masking-sample}D, showing overlaps and gaps, and some misidentifications caused by noise and boundary effects.
    Overlaps are resolved by assigning each contested point to the observation point closest to it.
    Gaps are resolved by the expansion method described in the text.
    (B) The depiction of the characteristic left, center, and right dominant directions of the $(V_{P_1}, V_{P_2})$ configuration space.
    (C) The ``filled-in'' map.
    The red, green, and blue colors indicate left, central, and right orientation, respectively.
    The gray color in (A) and (C) indicates the ND domain, while the black color indicates pixels that were not assigned to any polygon during the modeling phase.
    }
    \label{fig:tiling-process}
\end{figure*}

The polygon creation process produces a polygonal tiling of the binary transition mask.
Since the polygons are modeled independently, the resulting models may overlap partially, and various-sized gaps may occur between them at this stage, as clearly visible in Figure~\ref{fig:tiling-process}A. 
Such inaccuracies are most likely to occur in regions where the centralization algorithm is semi-stable, i.e., when transition lines become roughly parallel (the CD area), and in the transitional area at high noise levels. 
To fine-tune the proper tiling of the CSDs, we treat the polygon models themselves as data and use statistical methods to identify trends in them.
These trends, properly interpreted, constitute the domain decomposition of the CSD.

\subsection*{Statistical inference and domain reconstruction}
\label{ssec:stat-inf}
The polygon models provide an assemblage of discrete information about the underlying CSDs.
From this information, \textsc{GAnnET} gathers statistics, isolates relevant features of the binary transition mask, and identifies and filters noise that has come through the filtering mechanisms to this point.

The first relevant feature the \textsc{GAnnET} collects statistics on is the fit of the polygon models to their respective extended fingerprints.
Let $\mathcal{F}_{x_o}$ be an extended fingerprint from an observation point $x_o$, $A$ denote its set of terminal points, and $B$ denote the set of all points on its polygon model except those that lie on any aperture segments. 
The \textit{model error at an observation point $x_o$} is defined as $\mathcal{E}(x_o)=\rho_H(A,B)$.

After the polygon around each observation point $x_o$ receives the error score $\mathcal{E}(x_o)$, the scores are analyzed collectively to determine their statistical spread.
Empirically, low-noise scans have model errors tightly clustered near $0$ (see Figure~\ref{fig:model-building}C), while high-noise scans show large deviations (see Figure~\ref{fig:model-building}D).
If a polygon model's $\mathcal{E}(x_o)$ lies more than two standard deviations away from the mean, it is discarded.
This creates gaps in the tiling of the binary transition mask, which are overcome through statistical methods.

\subsubsection*{Establishing directionality}
After the polygons are created and filtered by fit quality, the resulting models are used to reproduce the gross characteristics of the binary transition mask.
In particular, polygon models are expected to cluster by their orientation into the scan's \textit{dominant directions}.
For a DQD, there are three dominant directions highlighted in Figure~\ref{fig:tiling-process}B with colored arrows: left (blue), right (red), and center (green) corresponding to the left, right, and combined DQD and central QD states, respectively.
We note that the dominant directions are not the same as the measurement direction.

\textsc{GAnnET} does not assume the existence of all directions in a given scan.
Rather, the algorithm is designed to automatically recognize which directions are present in a given CSD and which states need to be characterized.
Accordingly, the method does not require all canonical domains to be present in a CSD and can operate on scans containing only a subset of the expected states.
The dominant directions in the binary transition mask are determined by clustering the polygon \textit{orientations} defined by the unit normal to the median line through the model.
The clustering is carried out by the heat flow clustering algorithm (see \hyperref[sec:methods]{methods}).~\cite{Weber24-HFC}

\subsection*{Creating 4-state domain decomposition}
\label{sssec:4-state-lab}
The next step of \textsc{GAnnET} is to create the domain decomposition by identifying and grouping individual polygons.
The global state determination is performed in two phases.
First, \textsc{GAnnET} assigns each polygon to one of three orientation-based domains (left, central, and right).
Then, the central domain is split into DD and SD$_C$ states. 

The polygon models obtained at this point might overlap, and the tiling of the CSD might contain gaps, as depicted in Figure~\ref{fig:tiling-process}A.
Before global state labels can be assigned, overlaps must be resolved and any existing gaps filled, ensuring that each point in the scan is assigned to a unique polygon.
To eliminate the overlaps, \textsc{GAnnET} assigned each point claimed by two or more polygons to the nearest polygon center, as long as the direct line between the point and that center does not cross a transition line.
To remove the gaps between polygons, the territory each polygon inhabits is expanded until all gaps are filled.
The expansion process uses a standard method in which polygons are expanded pixel by pixel until they abut pixels claimed by another polygon.

\begin{figure*}[t]
    \centering
    \includegraphics{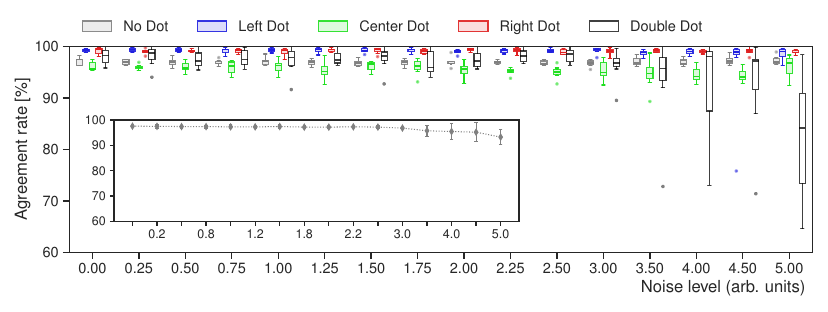}
    \caption{\textbf{The performance of GA\fakeSC{nn}ET on simulated CSDs with varied noise levels.}
    The noise levels used in all tests with simulated data range from $0.00$ to $5.00$, where the reference noise level extracted from the experimental data is denoted, for simplicity, as $1.00$. 
    The boxes span the interquartile range ([IQR], 25th--75th percentiles), the whiskers extend to 1.5 times the IQR from the box limits, and the horizontal lines inside boxes denote the median. 
    Points outside the whiskers represent outliers.
    The inset shows the overall performance across all domains and CSDs at each noise level.
    Data are presented as means, and error bars represent the sample standard deviation.
    }
    \label{fig:results-stats}
\end{figure*}

In the resulting filled-in map, each pixel is classified as belonging to the no-dot, left, central, or right orientation, though some polygons might, at this point, be misclassified (see Figure~\ref{fig:tiling-process}C). 
To correct potential misclassifications, \textsc{GAnnET} re-examines all polygons within each class. 
The underlying expectation is that the left\nobreakdash-, center\nobreakdash-, and right-class-assigned polygons should be mostly grouped contiguously on the left, center, and right sides of the scan, respectively.
Individually misclassified polygons will likely be haphazardly distributed throughout the scan.

The filled-in map is used to create the 4-state domain decomposition into ND, SD$_L$, SD$_C$, and SD$_R$ states.
The directionality re-examination is performed per class for left- and right-handed orientations.
For each class, a 0-or-1 map is created, with $1$ indicating a pixel attached to the class under consideration and $0$ indicating all alternative classes.
To identify the most probable region of the scan to assign as the relevant domain, the algorithm first constructs a convex hull for each contiguous grouping within a given class.
Then the area of each hull is computed, and the region with the largest area is selected.
Lastly, the selected convex hull is modeled as a separate polygon using the same fingerprinting techniques used to create the smaller polygon models.
This may result in assigning the fingerprint-modeled polygons to more than one domain.
Points that belong to such multiple-domain polygons are used to determine the gradual transition between global state domains.
This allows modeling the uncertainty typical of real-world scans, where domain boundaries are not sharp.

\subsection*{Creating 5-state domain decomposition}
\label{sssec:5-state-lab}
The center-orientation region must be divided into DD and SD$_C$ domains.
However, the polygon orientation is insufficient to differentiate between the SD$_C$ and DD states.
Rather, \textsc{GAnnET} employs physical principles to make this determination.
Since the transition between the SD$_C$ and DD states is not sharp, to label the scan, we must be able to measure a degree of grading between the states.

In the SD$_C$ state, the valence charge carriers are not localized on either QD but form a pair that acts as an SD.
In the DD state, in contrast, the charge carriers interact, but each is on a distinct QD.~\cite{Wiel02-DQD}
In CSDs, the geometry of the $(m,n)$ polygonal cell provides information about the effective electrostatic regime of the coupled-dot system.
The orientation and spacing of the cell edges contain a fingerprint of a QD device's capacitive environment, making these measurements useful tools for device characterization.~\cite{Losert26-AEC}

We interpret the charge-cell geometry within the constant-interaction picture for DQDs.~\cite{Beenakker91-COC, Wiel02-DQD}
In this framework, quadrilateral cells correspond to a regime with weak mutual coupling, while increasingly hexagonal cells reflect stronger interaction-induced distortion of the cell geometry. 
The hexagonal cell geometry indicates that the barrier is large enough for the charge carriers to be (mostly) spatially localized, yet small enough to allow tunneling.
Indistinct cells indicate that the charge carriers have merged into a charge-position multiplet: the barrier is so low that they are no longer spatially separated.

We emphasize, however, that the visibility and interpretation of particular transition features are not universal and may depend on device geometry, reservoir connectivity, tunnel rates, electrostatic coupling, and sensor placement. 
Accordingly, in \textsc{GAnnET}, we use the observed cell geometry as an effective indicator for grading between SD$_C$ and DD behavior within the measurement setting considered here, rather than as a universal signature of a specific tunneling regime (see \hyperref[sec:methods]{methods}).

\begin{figure*}[t]
    \centering
    \includegraphics{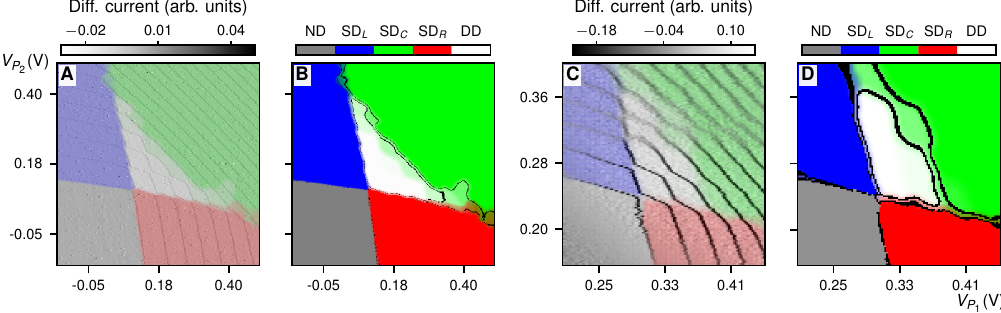}\\
    \caption{\textbf{Final probabilistic domain decomposition for simulated and experimental CSD.}
    (A) A sample noisy simulated CSD shown in Figure~\ref{fig:masking-sample}A and (C) a sample experimental scan shown in Figure~\ref{fig:masking-sample}D overlaying the charge sensor data.
    Regions with confidence in the dominant label of at least $70~\%$ are represented as ringed areas for (B) the simulated data and (D) the experimental scan. 
    }
    \label{fig:results-visuals}
\end{figure*}

\subsection*{Probabilistic state decomposition}
The final domain decomposition involves the assignment of a vector-valued function from the pixel location $V_{ij}=(x_i,y_j)$ within the CSD to a probability vector ${\bf p}(V_{ij})=(p_{\rm ND},\,p_{{\rm SD}_L},\,p_{{\rm SD}_C},\,p_{{\rm SD}_R},\,p_{\rm DD})$ such that each component on vector ${\bf p}(V_{ij})$ is non-negative and $p_{\rm ND}+p_{{\rm SD}_L}+p_{{\rm SD}_C}+p_{{\rm SD}_R}+p_{\rm DD}=1$.~\cite{Kalantre17-MLD}
The components of the vector ${\bf p}(V_{ij})$ represent the probability that the point $V_{ij}$ belongs to one of the five domains: $p_{\rm ND}$ is the probability that the point $V_{ij}$ is in the ND region; the $p_{{\rm SD}_L}$, $p_{{\rm SD}_C}$, and $p_{{\rm SD}_R}$ are probabilities that point $V_{ij}$ is in the left-, center-, and right-SD regions, respectively; and $p_{\rm DD}$ is the probability that point $V_{ij}$ is in the DD domain (see \hyperref[sec:methods]{methods}).

\subsection*{Technical Validation}
\label{ssec:val-results}
In simulated settings, noise and sensor artifacts can be modeled and controlled.~\cite {Ziegler22-TRA}
This enables systematic testing of \textsc{GAnnET} 's performance across varying levels of data quality degradation.
To validate \textsc{GAnnET}, we use a set of 7 qualitatively distinct simulated DQD devices with varying noise levels.
The noise levels used in these tests are varied around a reference level extracted from the experimental data, with the reference noise level denoted, for simplicity, as $1.0$.~\cite{Ziegler22-TAR, Ziegler23-AEC}

For each simulated CSD, we use 16 noise levels ranging from $0.0$ (noiseless data) to $5.0$. 
This results in $112$ distinct simulated CSDs.
In addition, we test the performance of the \textsc{GAnnET} on a set of 9 large experimental scans from the \textit{QFlow 2.0} dataset.~\cite{qf-data}
Because \textsc{GAnnET} is a deterministic, non-ML labeling procedure, the same algorithmic settings are applied uniformly across both validation sets.

\subsection*{Validation on simulated datasets}
Using the ground-truth labels assigned at the pixel level during simulations and the \textsc{GAnnET} labels, we can quantify both the overall per-CSD agreement between labels and the per-class performance. 
The ground-truth labels for the simulated data, $\mathcal{L}(V_{ij})$, are stored as one-hot vectors, with the hot component indicating the assigned state.
The \textsc{GAnnET} algorithm provides a map in which each pixel is assigned a probability vector ${\bf p}(V_{ij})$.

To quantify overall performance, we compare the two vectors at each point in each simulated CSD.  
Pixels for which \textsc{GAnnET} shows an agreement with ground truth of at least $70~\%$ are flagged as \textit{correct}.
Otherwise, they get an \textit{incorrect} flag.
We observe an overall state assignment agreement of around $97~\%$ when averaged over all CSDs for noise levels up to $3.0$, as shown in the inset of Figure~\ref{fig:results-stats}.
Once the noise level surpasses $3.0$, the agreement deteriorates slightly to around $(95.6 \pm 3.3)~\%$ at noise level $4.0$ and $(93.3\pm2.8)~\%$ at noise level $5.0$.
For reference, the performance of a recently proposed QD charge-tuning algorithm begins to decline at a noise level of $2.0$.~\cite {Ziegler22-TAR}

We also compute the proportion of correctly assigned pixels to the total number of pixels for each of the five domain regions, as determined by the ground-truth labels.
Figure~\ref{fig:results-stats} shows the performance of \textsc{GAnnET} on a per-domain basis for each noise level for the 7 simulated devices.
The ND, SD$_L$, and SD$_R$ state identifications remain almost perfectly robust up to the highest noise level considered.
The CD state shows a slight decline in performance at higher noise levels (around $3.0$).
The DD state assignment is the most fragile and breaks down once the simulated noise levels surpass $3.5$.

The statistical determination, error mitigation, and data redundancy ensure stable state assignments for ND, SD$_L$, and SD$_R$.
Since the domains in the CSD decomposition are complementary, the high, consistent performance in classifying these three domains naturally translates into comparable performance in determining the combined SD$_R$ and DD domains (the central orientation region).
\textsc{GAnnET}'s sub-statistical boundary determination between SD$_R$ and DD domains results in performance degradation at higher noise levels.

Figure~\ref{fig:results-visuals}A shows the final domain decomposition for a sample noisy simulated CSD shown in Figure~\ref{fig:masking-sample}A.
The boundary between the SD$_R$ and DD regions is determined by a series of measurements on individual polygons quantifying their \textit{hexagonness} (see Equations~\ref{eq:ul} and \ref{eqn:hexagonness} in \hyperref[sec:methods]{methods}).
Because this process relies on measurements taken on individual polygons, it lacks redundancy or statistical error checking, making the SD$_R$ and DD state assignments more sensitive to noise.
However, despite this limitation, the SD$_R$ and DD state assignments remain quite robust up to a fairly high level of noise, as can be seen in Figure~\ref{fig:results-stats}.
In Figure~\ref{fig:results-visuals}B, regions with confidence in the dominant label of at least $70~\%$ are represented as ringed areas.
The gradual change in the color indicates transitional labeling.

\subsection*{Validation on experimental datasets}
For the experimental data, ground-truth labels are unavailable. 
It is thus not possible to perform large-scale quantitative validation of \textsc{GAnnET} performance.
Thus, when assessing the algorithm's outputs for experimental data, we focus on the qualitative features resulting from the assigned labels.
In particular, we focus on the overall positions of the individual states within the 2D configuration space, such as the ND state in the bottom-left corner and the SD$_L$ and SD$_R$ states in the top-left and bottom-right corners, respectively.
We also consider the agreement between the charge transition line geometry (e.g., horizontal or vertical parallel lines) and the assigned global state label (e.g., SD$_R$ or SD$_L$). 
Finally, we look at the correctness of the transitional regions, i.e., the DD state blending into SD$_L$ as $V_{P_1}(V)$ decreases, into SD$_R$ as $V_{P_2}(V)$ decreases, and into SD$_C$ as $V_{P_1}(V)$ and $V_{P_2}(V)$ increase.

A sample probabilistic domain decomposition for the experimental scan shown in Figure~\ref{fig:masking-sample}D, indicating the probabilistic assignment of the five possible states, is depicted in Figure~\ref{fig:results-visuals}C.
Figure~\ref{fig:results-visuals}D highlights in the configuration space regions where the label confidence surpasses $70~\%$ as well as regions where the labels indicate a transition between states (i.e., all components of the probability vector ${\bf p}(V_i)$ are less than $0.7$). 
A visual inspection of this map confirms a high level of agreement between \textsc{GAnnET}'s assigned labels and the human interpretation.
To further validate the quality of the domain decomposition returned by \textsc{GAnnET}, we consulted with two external experts, one from academia and one from the industry.
Both experts are experimentalists with long experience in the semiconductor QD domain. 
The received feedback further confirms the high level of agreement between the automatically generated labels and the experts' manual labeling of the data.

\section*{Discussion}
\label{sec:discussion}
\textsc{GAnnET} is a critical component of a public-private-academic partnership to enable systematic data sharing in the semiconductor QD community.~\cite{Zwolak23-DNC}
The lack of robust data-sharing mechanisms has been widely recognized as a barrier to reproducibility and to the rapid iteration of automation methods.~\cite{Akrap26-RRC}
A comprehensive and reliable reference database of \emph{labeled experimental} measurements is essential for developing, validating, and benchmarking automation tools for QD characterization and control.
\textsc{GAnnET} provides a practical first step toward building such a database by enabling automated, physics- and geometry-informed annotation of CSDs at scale.

We demonstrated that \textsc{GAnnET} can systematically process experimentally acquired CSDs of varying quality and resolution, producing reproducible global state-domain labels.
Validation on simulated data shows that \textsc{GAnnET} remains reliable even at noise levels exceeding those typically observed in experiments, supporting its use as a labeling engine for large-scale dataset construction.
At present, \textsc{GAnnET} requires a binarized transition mask as input, and the thresholding step still requires modest human input to select gradient thresholds and filter parameters.
This preprocessing requirement is, therefore, the primary bottleneck to fully automated deployment.
We are actively developing a fully automated binarization module; however, progress is limited by the scarcity of publicly available moderate- and poor-quality experimental CSDs, as experimental datasets are still often released only upon ``reasonable request'' or not shared at all.~\cite{Zwolak23-DNC, Akrap26-RRC}

The broader aim of this effort is to establish concrete mutual benefits of data sharing.
\textsc{GAnnET} offers a value proposition by converting raw experimental scans into standardized, labeled data products suitable for method development and benchmarking.
Once the automated thresholding module is complete, we anticipate deploying a lightweight web-based interface for uploading and processing CSDs with \textsc{GAnnET}, with the resulting annotations and metadata systematically incorporated into the QFlow 2.0 data infrastructure.~\cite{qf-data}

Finally, as commercial offerings for QD tuning and calibration software increasingly incorporate ML and other AI components, independent and rigorous validation of such systems will soon become essential.
In such settings, performance claims can be difficult to interpret without standardized benchmarks, transparent evaluation protocols, and access to representative experimental data.
A curated reference database of labeled experimental CSDs\,---\,including data that may be subject to access controls---can provide an objective basis for evaluating both academic and commercial automation systems under realistic operating conditions.

Analogous dynamics have played out in large language models, where rapid commercialization has driven the emergence of standardized, continuously updated evaluation ecosystems that help separate marketing claims from reproducible performance comparisons.~\cite{lmsys-arena, helm}
In addition, several industry-facing evaluation and observability platforms provide ``behind-the-wall'' workflows for testing models against curated private datasets and controlled protocols.~\cite {langsmith-eval, phoenix}
These precedents motivate a similar validation pathway for QD autotuning systems: a trusted reference database of labeled experimental measurements, potentially with access controls, to enable rigorous evaluation under realistic operating conditions.
In this context, \textsc{GAnnET} can serve as part of a ``behind-the-wall'' labeling and validation pipeline for building and maintaining trusted reference datasets, while broader public releases can proceed in stages aligned with partnership agreements, data rights, and community norms.~\cite{Zwolak23-DNC}

\section*{Methods}
\label{sec:methods}
\subsection*{Data preprocessing}
\label{ssec:data-prep}
Raw CSDs (simulated or experimental) represent the device response evaluated on a 2D grid of gate-voltage configurations, where each pixel value corresponds to a charge-sensor (or transport) signal correlated with the total charge.
\textsc{GAnnET} does not operate directly on the raw signal.
Instead, it requires a binarized representation of charge-transition locations, which we refer to as a binary transition mask.
In the binary transition mask, pixels consistent with no change in charge configuration are labeled $0$, while pixels corresponding to a charge transition (i.e., where a charge carrier loads from or unloads to a reservoir or transfers between QDs) are labeled $1$.
Examples of binary transition masks for the simulated and experimental CSDs in Figures~\ref{fig:masking-sample}A and \ref{fig:masking-sample}D are shown in Figures~\ref{fig:masking-sample}D and \ref{fig:masking-sample}F, respectively.

In practice, both simulated and experimental CSDs can be noisy and may contain asymmetries and artifacts arising from measurement electronics, drift, and device-specific features.
Consequently, denoising and binarization currently require modest human input, primarily in selecting gradient thresholds and filter parameters.
For high-noise data, additional correlation-based strategies are applied to reduce spurious detections and broken transitions.
All preprocessing steps that produce the binary transition mask are performed outside of \textsc{GAnnET} in the present implementation.
This preprocessing requirement currently accounts for the majority of manual labor in the workflow; however, once a binary transition mask is provided, all subsequent stages of \textsc{GAnnET} operate fully automatically.

\subsection*{Normalized Hausdorff distance}
\label{ssec:H-dist}
Let $A$ and $B$ be nonempty, finite sets of points within a metric space with metric $\rho$.
The normalized $\ell^2$ Hausdorff distance is
\begin{equation}
    \rho_H(A,\,B) \;=\;\left(\textstyle\frac12\mathcal{D}(A;B)^2\,+\,\frac12\mathcal{D}(B;A)^2\right)^{\frac12}, \label{eqn:HausdorffDef}
\end{equation}
where $\mathcal{D}(A;B)$ is the $\ell^2$-deviation of $A$ with respect to $B$ defined as 
\begin{equation}
    \mathcal{D}(A;B) =
    \bigg(\frac{1}{|A|\,\text{diam}(A)^2}\sum_{x\in{A}}\text{dist}(x,B)^2\bigg)^{\frac12},
\end{equation}
where $\text{dist}(x,B)=\min\left\{\rho(x,\,y)\;\big|\;y\in{}B\right\}$, $|A|$ denotes the cardinality of $A$, and $\text{diam}(A)=\sup\left\{\text{dist}(x,\,y)\;\big|\;x,y\in{}A\right\}$ denotes the diameter of set $A$.

The set function $\mathcal{D}$ is not symmetric in $A$ and $B$.
Intuitively, $\mathcal{D}$ measures how far, in a root-mean-square sense, a typical point of $A$ lies from the nearest point of $B$.
For example, $A\subseteq{}B$ implies $\mathcal{D}(A;B)=0$ but says nothing about $\mathcal{D}(B;A)$.
$\mathcal{D}$ is unitless and scale invariant.
The set function $\rho_H$, on the other hand, is symmetric in $A$ and $B$ and $\rho_H(A,B)=0$ if and only if $A$ and $B$ are the same set.

\subsection*{Polygon fitting}
\label{ssec:poly-fit}
If the polygon has $n$ vertices (with $n=4,5,6$ for the DQD system), then $\rho_H$ must be minimized in a $2n$-dimensional configuration space.
This search is carried out in two steps: a rough search followed by a fine search.
In the rough search phase, all terminal points for a given extended fingerprint are ordered sequentially and each possible subset of $n$ sequential points is used to build a polygon model.
We then find the smallest value of $\rho_H$ among the resulting models.
For $N$ terminal points, this means testing ${{N}\choose{n}}=\frac{N!}{(N-n)!n!}$ possible models.
When $n$ is small and $N\gg n$, this is larger than $(N-n)^n/n!$, which can become computationally prohibitive for large $N$.
 
To reduce the number of models that need to be checked, we employ several mitigation strategies.
For quadrilateral models, we reduce the number of models to test to about $2\,(N/2)^2$ by requiring that two of their vertices be diameter points of the fingerprint.
When testing pentagonal or hexagonal models, we first search for gaps at the termination points and then force the model apertures to span one or more of them.
If the number of gaps is $M$, this reduces the search to $M\cdot{}O(N^2)$ models in the pentagonal case and ${{M}\choose{2}}\cdot{}O(N^2)$ in the hexagonal case.

The second phase, the fine search, begins with the model selected in the rough search.
Because the search space in the rough-fitting phase was restricted to terminal points, it is possible that nearby points not included in the extended fingerprint could yield a substantially better fit.
As noted earlier, $\rho_H$ can be considered a function on a $2n$-dimensional space.
To optimize the fit, we employ the discrete gradient flow method on the gradient of $\rho_H$ in this space.
To prevent oscillatory behavior, the model's vertex points can move at most one pixel from their previous location at each step in the flow.
The number of steps necessary for convergence depends on the internal characteristics of the extended fingerprint, e.g., the size of the largest gaps between terminal points.
In practice, we observe convergence in no more than 6 steps, except for polygons in very noisy regions, where this number can increase.
The fine search is computationally less demanding than the rough search, as gradients can be computed in polynomial time with respect to the number of polygon parameters.
In our case, the computation cost scales as $N\cdot{}O(n)$ where $N$ is the number of fingerprint terminal points.
Since in our application $n\ll N$, this is better than the $O(N^2)$ computation time for the rough search.
The result of the fine search is a polygon model for each extended fingerprint that locally minimizes the $\rho_H$ cost function.

\subsection*{Heat flow clustering algorithm}
\label{ssec:hfc-alg}
\begin{figure}[t]
    \includegraphics[width=\linewidth]{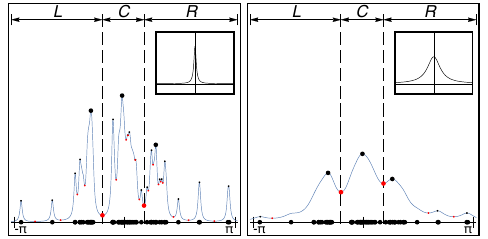}
    \caption{
    \textbf{Two kernel choices in the heat flow clustering method.}
    The $x$ axis parameterizes $\mathbb{S}^1$, and the points are polygon orientations.
    The inset in each plot shows the kernel choice.
    Properly selected kernels can lead to clear, persistent clustering of points, with each point ultimately assigned to the cluster in which it appears most persistently.
    \emph{L}, \emph{C}, and \emph{R} indicate left, center, and right orientation clusters, respectively.
    }
    \label{fig:heat-flow}
\end{figure}

The heat flow clustering algorithm is a modification of the fairly common potential-based clustering methods. 
It works by selecting a static kernel $k(x)$ that resembles a Gaussian, and then scaling parabolically to obtain the time-dependent function
\begin{equation}
    K(x,t)=k(x/\sqrt{t})/\sqrt{t}.
\end{equation}
A selection of discrete times $\{t_1,\dots,t_N\}$ is chosen, and a clustering is performed at each time $t_i$, with $N=15$.
Final cluster assignments are determined by assigning each point to the cluster to which it most persistently belongs.
Figure~\ref{fig:heat-flow} shows two kernel choices in the heat flow clustering method.
The heat flow clustering algorithm yields as few as one and as many as three cluster points, which correspond to the one, two, or three characteristic directions.

\subsection*{Measuring hexagonness of polygon cells}
\begin{figure}[t]
    \centering
    \includegraphics{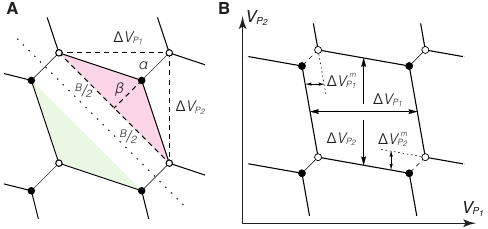}
    \caption{\textbf{Measuring the \textit{hexagonness} of cells.}
    (A) The division of the hexagonal cell into an upper (above the centerline) and lower (below the centerline) polygon showing the cell roof (pink) and floor (green).
    (B) The polygonal cell's dimensions in terms of physical quantities in the constant-interaction model.~\cite{Wiel02-DQD}
    }
    \label{fig:hexagon-model}
\end{figure}

To create a physically meaningful quantitative measure of the \textit{hexagonness} of a cell, we divide each cell along its centerline (the dotted line in Figure~\ref{fig:hexagon-model}A) into an upper and lower cell (the regions above and below the dotted line, respectively). 
Then, we define the cell \textit{roof} as the region above the dashed line of length $B$ in the upper cell in Figure~\ref{fig:hexagon-model}A and the cell \textit{floor} as a corresponding region in the lower cell.
After the polygon models have been built, the following ratios are easily calculated:
\begin{equation}
    U\;=\;\frac{\mathrm{A(roof)}}{\mathrm{A(upper\;cell)}}
    \quad\mathrm{and}\quad
    L\;=\;\frac{\mathrm{A(floor)}}{\mathrm{A(lower\;cell)}}.
\end{equation}
We interpret these geometric ratios in terms of the constant interaction model, with $B=[(\Delta{}V_{P_1})^2+(\Delta{}V_{P_2})^2]^{\frac12}$, $\alpha=\frac12[(\Delta{}V_{P_1}^m)^2+(\Delta{}V_{P_2}^m)^2]^{\frac12}$, $\beta=-\alpha+\frac{\Delta{}V_{P_1}\,\Delta{}V_{P_2}}{B}$, and $\alpha,\beta\in[0,\frac12B]$, where $\Delta{}V_{P_i}$ and $\Delta{}V_{P_i}^m$ denote the spacing between the charge transitions (see Figure~\ref{fig:hexagon-model}B). 

While in the idealized constant-interaction model $U=L$, in real-world CSDs, these ratios are equal only if device characteristics, such as crosstalk and stray capacitances, vary negligibly within the dynamic ranges of the plunger voltages.
As this is not true in general, the values of $U$ and $L$ are expected to be different and are computed separately as 
\begin{equation}
    U,\,L\;=\;\frac{\frac12\beta{}B}{\frac12\beta{}B+\frac12\alpha{}B}
    \label{eq:ul}
\end{equation}

In practice, the values of $\Delta{}V_{P_1}^m$, $\Delta{}V_{P_2}^m$ change with the absolute values of the gate voltages $V_{P_1}$ and $V_{P_2}$ and can vary substantially even within the same hexagon giving $U$ and $L$ measurably different values.
The hexagonness measures $U$ and $L$ have a concrete physical interpretation, which is easiest to explain for the idealized, symmetric case where $\Delta{}V_{P_1}\approx\Delta{}V_{P_2}\equiv\Delta{}V_P$ and $\Delta{}V_{P_1}^m\approx\Delta{}V_{P_2}^m\equiv\Delta{}V_P^m$. These symmetries allow to simplify Equation~\ref{eq:ul} as 
\begin{align*} 
    U\;\approx\;1-\frac{\Delta{}V_P^m}{\Delta{}V_P}
    \;\approx\;1-\frac{C_m}{C} \\ 
    L\;\approx\;1-\frac{\Delta{}V_P^m}{\Delta{}V_P}
    \;\approx\;1-\frac{C_m}{C},
\end{align*}
where $C_m$ and $C$ refer to the charge coupling constants from the constant-interaction model.~\cite{Wiel02-DQD}
The range is $\Delta{}V_P^m\in[0,\Delta{}V_g]$ so that $U,L\in[0,1]$ with $0$ indicating a completely joined SD$_C$, $1$ indicating fully decoupled DD, and $U,L\in(0,1)$ indicating intermediate states of electron-electron coupling.

Although quantitative grading between DD and SD$_C$ states can be inferred geometrically from the CSD, appropriate cutoffs must be chosen to assign each polygon a specific label.
From the area ratio, we create the quantitative hexagon-ness score $\mathcal{H}$ for each polygon
\begin{equation}
    \mathcal{H}\;=\;
    \begin{cases}
        \frac{1}{0.8}(U+L), & \text{if $0\le\frac12(U+L)\le0.4$} \\
        1, & \text{if $0.4<\frac12(U+L)\le1$}
    \end{cases} \label{eqn:hexagonness}
\end{equation}
The $\mathcal{H}$ can be interpreted as the probability that a given polygon represents a DD versus SD$_C$ state.
In practice, where these cutoffs are set will depend, to some extent, on the application.
The cutoffs are user-defined decision thresholds rather than learned parameters, and in the validation presented here, they are held fixed across all datasets.
The \textsc{GAnnET} is fully customizable and can be adjusted to meet research needs.

\subsection*{Probability grading between the adjacent domains}
\label{ssec:probs_deriv}
The probabilities are assigned in three distinct phases.
The polygonal modeling described in the \hyperref[sssec:4-state-lab]{Creating 4-state domain decomposition} section results in assigning most pixels to one of the ND, SD$_L$, SD$_C$, or SD$_R$ state domains (DD assignments not created yet).
However, the small fingerprint-defined polygons lying near the domain boundaries might overlap with two or more state domains.
When assigning probability vectors, points inside polygons that are partially absorbed by an adjacent domain are reassigned to the original class based on the region where the center of the polygon lies.
In other words, the initial 4-domain decomposition assigns to each point a state vector
\begin{equation}
    {\bf p}^{(4)}(V_i)=(p_{\rm ND},\,p_{{\rm SD}_L},\,p_{{\rm SD}_C},\,p_{{\rm SD}_R},\,0),
\end{equation}
where exactly one of $p_{\rm ND}$, $p_{{\rm SD}_L}$, $p_{{\rm SD}_C}$, and $p_{{\rm SD}_R}$ is $1$ and the others four components are $0$.

The second stage, described in section \hyperref[sssec:5-state-lab]{Creating 5-state domain decomposition}, alters the polygons in the central orientation region by assigning to each polygon its DD probability.
Using the score $\mathcal{H}$ given in Equation~\ref{eqn:hexagonness}, the state vector for points within this region is updated as 
\begin{equation}
    {\bf p}^{(5)}(V_i)=(0,0,\,1-\mathcal{H},0,\,\mathcal{H}).
\end{equation}
Outside of this region the state vector ${\bf p}^{(5)}(V_i)={\bf p}^{(4)}(V_i)$, with exactly one of $p_{\rm ND}$, $p_{{\rm SD}_L}$, and $p_{{\rm SD}_R}$ equal $1$ and the others components equal to $0$.

Finally, in the third stage, the state vectors are probabilistically interpolated between the adjacent state domains.
The assignment of individual polygons to each state makes an artificially sharp and rigid boundary between the no-dot, left, central, and right orientation-based domains (see Figure~\ref{fig:tiling-process}C).
At the same time, just as there is no completely sharp boundary between the SD$_C$ and DD state domains, there is no completely sharp distinction between the SD$_C$ and SD$_L$ or between the SD$_C$ and SD$_R$.

The probability grading between the outer SD$_L$ and SD$_R$ states and the inner SD$_C$ and DD states is determined geometrically.
The individually labeled polygons intrude into neighboring regions, as shown in Figure~\ref{fig:tiling-process}C.
We measure how far polygons from one state extend into the regions of another state, and take these distances as the standard deviations of the uncertainties in the region boundaries.
We then convolve the 5-vector with a local kernel of width equal to this deviation.
The result is a probability vector ${\bf p}(V_{ij})$ that grades among the five regions based on the uncertainty in the boundary location.

\section*{Resource Availability}
\subsection*{Lead contact}
\vspace{-2pt}
Requests for further information should be directed to and will be fulfilled by the lead contact, Justyna P. Zwolak (\href{mailto:jpzwolak@nist.gov}{jpzwolak@nist.gov}).

\subsection*{Materials availability}
\vspace{-2pt}
This study did not generate new materials

\subsection*{Data and code availability}
\vspace{-2pt}
This paper analyzes existing, publicly available data, accessible at \href{https://doi.org/10.18434/T4/1423788}{doi:10.18434/T4/1423788}.
Any additional information required to reanalyze the data reported in this paper is available from the lead contact upon request.

\section*{Acknowledgment}
\vspace{-2pt}
The authors acknowledge A. Chatterjee and F. Luthi for providing feedback about the automatically labeled experimental data.
The views and conclusions contained in this paper are those of the authors and should not be interpreted as representing the official policies, either expressed or implied, of the U.S. Government. 
The U.S. Government is authorized to reproduce and distribute reprints for government purposes, notwithstanding any copyright noted herein. 
Any mention of commercial products is for information only; it does not imply recommendation or endorsement by the NIST.

\section*{Author contributions}
\vspace{-2pt}
Conceptualization: JPZ conceived the fingerprinting-based labeling approach defined the system architecture, and the application objectives;
methodology: BW and JPZ jointly developed the auto-annotating framework and the automated labeling strategy;
software: BW developed and implemented the modeling, statistical inference, and re-modeling framework;
formal analysis: BW and JPZ jointly analyzed the data;
writing – original draft: BW and JPZ wrote the paper.

\section*{Declaration of interest}
\vspace{-2pt}
JPZ and BW are inventors on a patent application related to this work (U.S. Patent Application serial number US 2026/0023910 A1 published on January 22, 2026).
BW is the founder and a director of Intelligent Geometries, LLC.
JPZ is an advisory board member of \textit{Newton}.

\section*{Declaration of Generative AI and AI-assisted technologies}
\vspace{-2pt}
During the preparation of this work, the author(s) used Grammarly Pro to improve grammatical accuracy. 
After using this tool, the authors reviewed and edited the content as needed and take full responsibility for the content of the publication.

%

\clearpage
\includepdf[pages={1}]{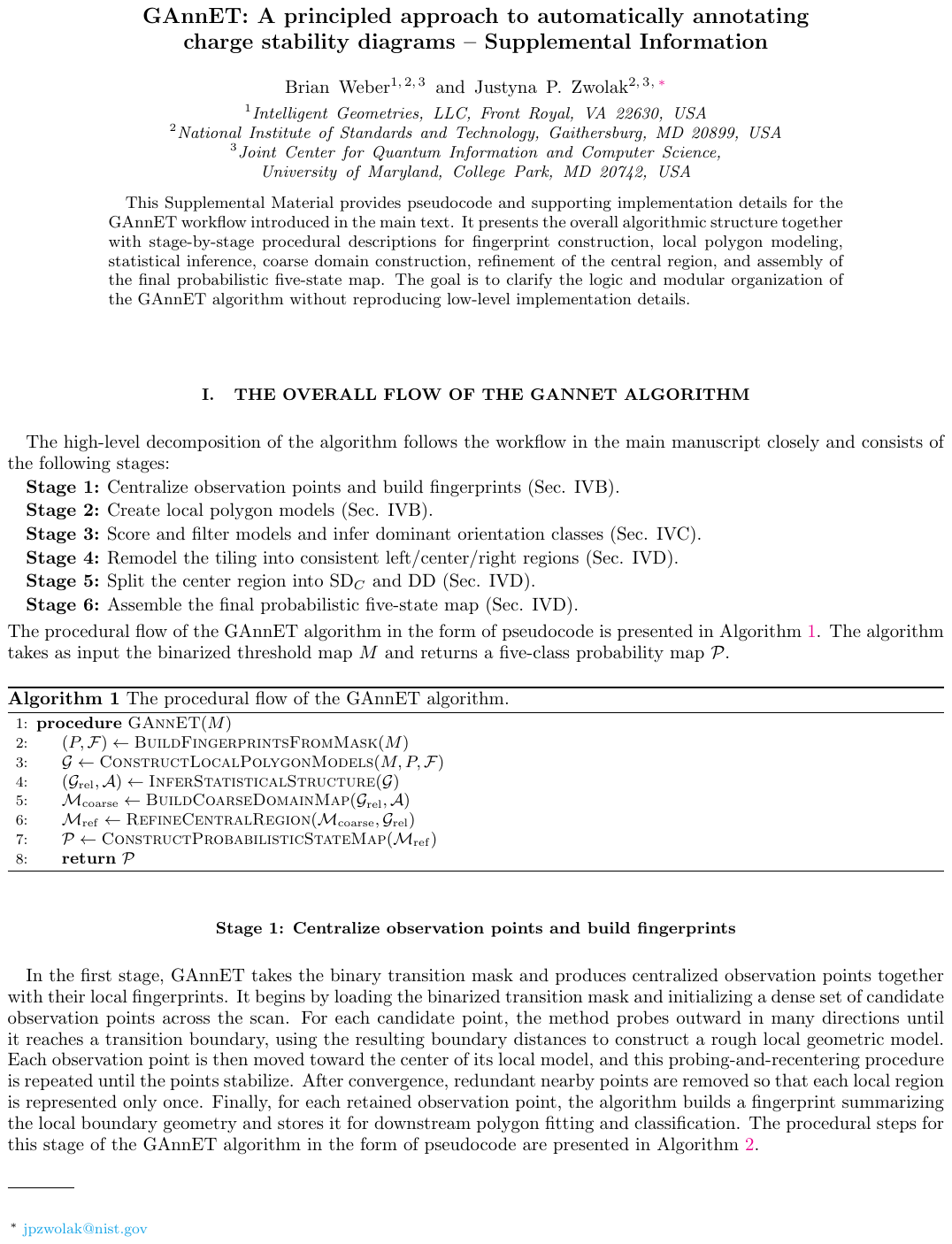}

\clearpage
\includepdf[pages={2}]{SM.pdf}

\clearpage
\includepdf[pages={3}]{SM.pdf}

\clearpage
\includepdf[pages={4}]{SM.pdf}
\end{document}